\documentclass[prl,twocolumn,showpacs]{revtex4}

\usepackage{graphicx}

\newcommand{\up}{\uparrow}
\newcommand{\dn}{\downarrow}

\newcommand{\bsigma}{{\mbox{\boldmath $\sigma$}}}

\newcommand{\ket}[1]{\mbox{$|#1\;\!\rangle$}}

\begin{document}

\title{Universal set of quantum gates for double-dot spin qubits with fixed interdot coupling}

\author{Ronald Hanson}
\affiliation{Center for Spintronics and Quantum Computation, University of California, Santa Barbara, California 93106, USA}

\author{Guido Burkard}
\affiliation{Department of Physics and Astronomy, University of Basel, Klingelbergstrasse 82, CH-4056 Basel, Switzerland}


\begin{abstract}
We propose a set of universal gate operations for the singlet-triplet qubit
realized by two electron spins in a double quantum dot,
in the presence of a fixed inhomogeneous magnetic field.
All gate operations are achieved by switching the potential offset between the two dots with an electrical bias, and do not require time-dependent control of the tunnel coupling between the dots.
We analyze the two-electron dynamics and calculate the effective qubit rotation angle as a function of the applied electric bias.
We present explicit gate sequences for single-qubit rotations about
two orthogonal axes, 
and a CNOT gate sequence, completing the universal gate set. 
\end{abstract}


%
%

\maketitle

Electron spins in semiconductor quantum dots (QDs) are promising candidates for encoding and manipulating quantum information in the solid state. Initialization, manipulation and readout of electron spins have already been demonstrated in these systems~\cite{Petta,DelftRO}. Proposals exist for encoding one logical qubit in one~\cite{LD}, two~\cite{Levy,WuLidar,ByrdLidar}, three \cite{DiVincenzo,Kyriakidis2005}, or even more~\cite{Meier} spins. Although they differ in many respects, a common essential ingredient of all these proposals is electrical control of the two-electron exchange interaction in a double quantum dot, which is characterized by the singlet-triplet energy splitting $J$.

Conventionally, control over $J$ is envisioned through voltage control of the tunnel coupling $t$ between the two dots. However, in many QD systems, such as vertical pillars~\cite{Ono2002}, self-assembled dots~\cite{SAdots}, nanowires~\cite{NWs}, or etched dots in Si~\cite{SiDots}, $t$ is fixed by growth or fabrication parameters. Even for double QDs (DQDs) in electrically gated systems, such as GaAs dots~\cite{JeroPRB2003}
and carbon nanotubes~\cite{Mason}, fast control over the tunnel coupling is challenging and has not been demonstrated thus far.

A possible way around this problem was demonstrated in a recent experiment by Petta~\textit{et al.}~\cite{Petta}, where $J$ is controlled by the misalignment $\varepsilon$ between the two QDs. In contrast to the tunnel coupling, the misalignment can easily be changed over a wide range on a subnanosecond timescale by pulsing the source-drain bias~\cite{HayashiPRL2003} or a gate voltage~\cite{Petta}. Building on this result, Taylor \textit{et al.}~\cite{Taylor} proposed a set of universal gates for a logical qubit whose basis states are the two-electron states $|S\rangle=(\ket{\up\dn}-\ket{\dn\up})/\sqrt{2}$ and $|T_0\rangle=(\ket{\up\dn}+\ket{\dn\up})/\sqrt{2}$. However, their scheme requires $J$ to be tunable to zero, which is not possible by changing $\varepsilon$ alone~\cite{footnote1}. Therefore, voltage control of $t$ is still needed in their scheme.

Here, we propose a set of universal quantum gates for the $S-T_0$ qubit in a constant small inhomogeneous field, that eliminates the need for controlling the tunnel coupling $t$. We demonstrate how arbitrary single-qubit rotations can be performed at \textit{finite} $J$, by combining Z rotations with rotations around an axis in the XZ plane. We discuss the experimental requirements for this scheme and compare them to current-day devices. Finally, we outline a two-qubit CNOT operation, which is based on a change in the rotation angle of the target qubit that is conditional on the control qubit through spin-dependent tunneling and the capacitive coupling between qubits.

\textit{Tunable spin dynamics in a DQD}--%
Our qubit is realized in the $\ket{0}\equiv\ket{S}$ and $\ket{1}\equiv\ket{T_0}$ states of two electrons in a double quantum dot, where $S$ and $T_0$ are the lowest-energy singlet and triplet states.
The dynamics of these states can be described by the Hamiltonian
\begin{equation}
H = \left(\begin{array}{c c c c}
0 & \delta h/2 & 0 & 0\\
\delta h/2 & 0 & \sqrt{2} t & \sqrt{2} t\\
0 & \sqrt{2}t & U-\varepsilon & 0 \\
0 & \sqrt{2}t & 0   & U+\varepsilon
\end{array}\right),
\label{Hamiltonian}
\end{equation}
in the basis $|T_0\rangle$, 
$|(1,1)S\rangle$, $|(0,2)S\rangle$, $|(2,0)S\rangle$~\cite{footnote-triplets}. Here, ($m,n$) denotes the number $m$ ($n$) of electrons in dot $1$ ($2$), $\delta h$ is the inhomogeneous magnetic field between the dots and $U$ is the difference in Coulomb energy between the $(1,1)S$ and the $(0,2)S$ or $(2,0)S$ state. 

\begin{figure}
  \centerline{\includegraphics[width=6cm]{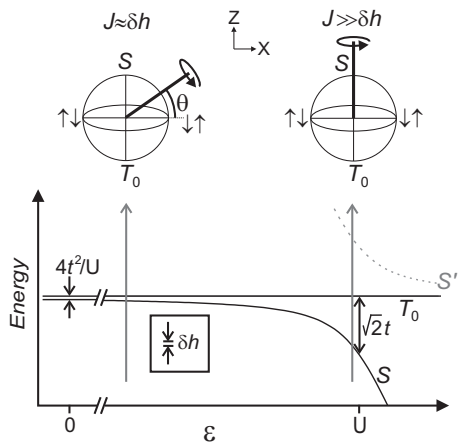}}
\caption{
Dependence of the two-electron energy levels in a double dot on the bias $\varepsilon$ for $\delta h$=0. Here, $S'$ denotes the first excited singlet state. In the presence of a small fixed inhomogeneous field $\delta h$ (see inset), the precession axis depends on $\varepsilon$, as illustrated by the Bloch spheres.
\label{fig:fig1}}
\end{figure}

Figure~\ref{fig:fig1} shows the energy of the lowest eigenstates as a function of $\varepsilon$ for $\delta h$=0. In this case, the eigenstates are pure spin states for all values of $\varepsilon$
and we define the qubit basis states
as the triplet $T_0$ and the lowest-energy singlet $S$, which are separated by an energy $J$. 
At $\varepsilon=U$, there is an avoided crossing of the $(1,1)S$ and $(0,2)S$ states, and as a consequence $J$ is large at this point.  In the presence of an inhomogeneous field $\delta h$ with magnitude much smaller than $t$ (as in the inset of Fig.~\ref{fig:fig1}), $S$ and $T_0$ remain eigenstates near $\varepsilon=U$ where $J\approx t\gg \delta h$. Far away from the avoided crossing, however, $J\approx \delta h$ and therefore $S$ and $T_0$ are strongly mixed. As a consequence, the qubit rotates about an axis determined by $\delta h$ and $J$.

The qubit subspace is energetically separated from the two remaining
singlets (away from the avoided crossing, the gap is $\approx U$).
Therefore, under the condition that $\varepsilon$ is always changed adiabatically with respect to the energy difference between the qubit states and all other states, the Hamiltonian (\ref{Hamiltonian}) can be reduced to the
qubit subspace and, in the qubit basis $\ket{S}$ and 
$\ket{T_0}$, has the general form
\begin{equation}
H = \frac{1}{2}\left(\begin{array}{c c}
J & \tilde{\delta h}\\
\tilde{\delta h}^* & -J
\end{array}\right) 
\equiv {\bf B}\cdot \bsigma,
\label{Heff}
\end{equation}
where we have chosen the zero of energy mid-way between the states 
$|0\rangle$ and $|1\rangle$ and introduced a pseudo-spin notation with 
Pauli matrices $\bsigma=(\sigma_x,\sigma_y,\sigma_z)$ in the
two-dimensional qubit subspace.
The pseudo-magnetic field is 
${\bf B}=({\rm Re}[\tilde{\delta h}],{\rm Im}[\tilde{\delta h}],J)/2$,
where the exchange coupling $J$ and effective difference field $\tilde{\delta h}$
are functions of $t$, $U$, $\delta h$, and $\varepsilon$.
In what follows, $\tilde{\delta h}$ will be real and thus the pseudo-field
always lies in the $XZ$-plane (Fig.~\ref{fig:fig1}). The angle
of the pseudo-field with the $X$ axis is
\begin{equation}
  \label{angle}
  \theta  = {\rm arctan}( J / \tilde{\delta h} ),
\end{equation}
and can be controlled by changing the electric bias $\varepsilon$,
while keeping $U$, $t$, and $\delta h$ fixed.
Single-qubit rotations can be carried out by switching between
two different values of the electric bias $\varepsilon$, as shown 
in Fig.~\ref{fig:fig1}.
One of these working points is chosen to lie close to the avoided crossing,
$|U-\varepsilon| \ll t$, where $J\approx \sqrt{2}t \gg \delta h$
(Fig.~\ref{fig:fig1}, right). 
At this point, ${\bf B}$ points into the $Z$ direction on the Bloch 
sphere and has a magnitude $B=J$.
The other working point is chosen far
from the avoided crossing, $|U-\varepsilon| \gg t$
(Fig.~\ref{fig:fig1}, left)
where the pseudo-field
${\bf B}$ lies close to the $X$ axis and $B=\sqrt{J^2+\tilde{\delta h}^2}$.
In theory, ${\bf B}$ can be made to align with $X$ by switching $t$ to zero. However, as we wish not to rely on this fast control of $t$, we assume $t$ is fixed and therefore $J$ remains finite. Thus, we cannot
reach a point where ${\bf B}$ lies in the equator plane, i.e., we have to work with a finite angle $\theta>0$.

\begin{figure}[t]
  \centerline{\includegraphics[width=7cm]{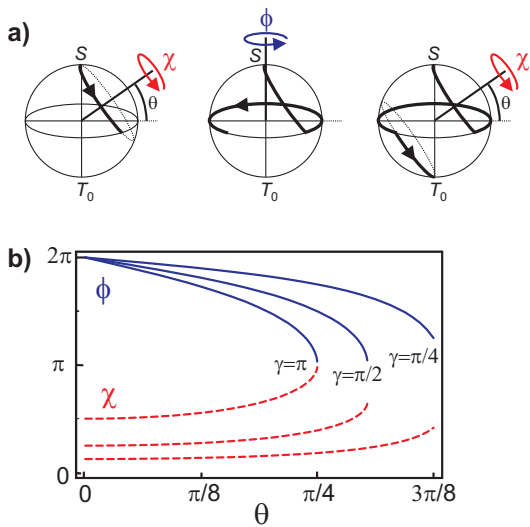}}
\caption{
\textbf{(a)} The three-step sequence Eq.~(\ref{x-sequence}) for 
X-rotations on the Bloch sphere.
\textbf{(b)} Rotation angles $\chi$, $\phi$ as functions of $\theta$,
producing rotations about $X$ by $\gamma=\pi,\pi/2,\pi/4$.
\label{fig:fig2}}
\end{figure}

\textit{Single-qubit gates}--%
We now show that arbitrary single-qubit 
rotations can be constructed from
the two available elementary operations: 
(i) rotations about the $\theta$-tilted
axis by some angle $\chi = B \tau$, $\tau$ being the switching time,
\begin{equation}
U_\theta(\chi) = \left(\begin{array}{c c}
\cos\frac{\chi}{2}+i\sin\frac{\chi}{2}\sin\theta & i\sin\frac{\chi}{2}\cos\theta \\
i\sin\frac{\chi}{2}\cos\theta   &   \cos\frac{\chi}{2}-i\sin\frac{\chi}{2}\sin\theta 
\end{array}\right),
\label{theta-rot}
\end{equation}
and (ii) nearly perfect $Z$ rotations by $\phi=J\tau\simeq \sqrt{2}t\tau$,
given by the diagonal matrix $U_Z(\phi)$ with diagonal entries
$e^{i\phi/2}$ and $e^{-i\phi/2}$.
%
Arbitrary single-qubit rotations can be constructed using the
Euler angle method, if rotations by arbitrary angles
about two orthogonal axes are available.  Therefore, it
is sufficient to show that arbitrary rotations about the $X$
axis, $U_X(\gamma)\equiv \exp(i\gamma\sigma_x/2)$, 
in addition to the $Z$-rotations, are feasible.
The three-step sequence
\begin{equation}
U_X(\gamma) 
=U_\theta(\chi) U_Z(\phi) U_\theta(\chi),
\label{x-sequence}
\end{equation}
with the rotation angles \cite{footnote-solution}
\begin{eqnarray}
\chi &=& {\rm arccos}
\frac
{\cos \frac{\gamma}{2} \sqrt{1-\tan^2\theta\sin^2\frac{\gamma}{2}}
-\sin^2\theta \sin^2\frac{\gamma}{2}}
{\cos^2\frac{\gamma}{2}+\cos^2\theta\sin^2\frac{\gamma}{2}},
\quad\label{chi}\\
\phi &=& -2\,{\rm arctan}
\frac
{\sin\chi \sin\theta}
{\cos^2\frac{\chi}{2}+\cos 2\theta\sin^2\frac{\chi}{2}},
\label{phi}
\end{eqnarray}
generates a rotation about the $X$-axis by an arbitrary angle $\gamma$,
as long as $0\le\theta\le\pi/4$.
One can intuitively understand this sequence by following the state on the Bloch sphere; Fig.~\ref{fig:fig2}a depicts the three steps for a rotation from \ket{S} to \ket{T_0} ($\gamma=\pi$). We note that switching between the working points has to be performed non-adiabatically with respect to $J$.
The rotation angles $\chi$ and $\phi$ are plotted as a function of 
$\theta$ in Fig.~\ref{fig:fig2}b for three different $X$-rotation
angles $\gamma$.
For a $\pi$-flip about the $X$ axis, $\gamma=\pi$,
we find the simpler expressions,
\begin{equation}
\chi = {\rm arccos}(-\tan^2\theta),\quad\quad
\phi = -2{\rm arctan}\frac{\sin\theta}{\sqrt{\cos 2\theta}}.
\label{chiphi-pi}
\end{equation}
We note that the sequence Eq.~(\ref{x-sequence}) is not simply 
one of the known NMR sequences. Actually, in NMR it is usually not
a problem to perform rotations about an axis in the equator plane
of the Bloch sphere \cite{Vandersypen}.

\textit{Doing nothing}-- 
A convenient ``idle'' position would be close to the avoided crossings $\varepsilon=\pm U$, i.e. close to the Z-gate operation point, as here only Z-rotations need to be accounted for. A disadvantage of this position is that the qubit is more susceptible to decoherence from charge fluctuations, due to the different orbital characters of the basis states close to the avoided crossing~\cite{HuDasSarmaPRL06}.
The best waiting position in terms of coherence is probably the 
symmetric point $\varepsilon=0$. However, since $\delta h$ and $J$ are of the same order at  $\varepsilon=0$, the spin constantly rotates about the pseudo-field ${\bf B}$ oriented between the $X$ and $Z$ axes at the angle $\theta$ from the $X$ axis.
To erase this effect, one could always wait for an integer number $n$ of full periods, $\tau=2\pi n/B$. Alternatively, a pulse sequence similar to refocusing in NMR~\cite{Vandersypen} can be applied,
\begin{equation}
  \label{sequence-idle}
  \openone = U_\theta(\chi) U_Z(\phi) U_\theta(\chi) U_Z(\phi),
\end{equation}
where $\theta$ and $\chi$ are angles determined
by the waiting position and time, and $\phi$ follows
from $\theta$ and $\chi$ as
\begin{equation}
  \label{phi-idle}
  \phi = {\rm arccos}
\left(1- \frac{2\cos^2 \frac{\chi}{2}
}{1-\cos^2\theta\sin^2\frac{\chi}{2}}
\right).
\end{equation}

\begin{figure}[b]
 \centerline{\includegraphics[width=8.5cm]{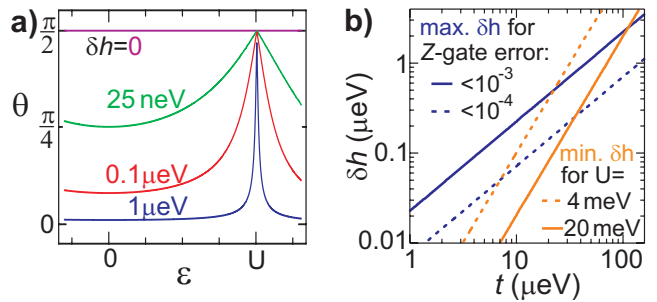}}
\caption{
\textbf{(a)} Angle $\theta$ as a function of $\varepsilon$ for different values of $\delta h$. Here, $U=4\,{\rm meV}$ and $t=5\,\mu{\rm eV}$.
\textbf{(b)} Blue lines: maximum value of $\delta h$ as a function of $t$, for different error thresholds for $Z$ rotations.
Orange lines:  minimum value of $\delta h$ as needed for $X$ rotation, for different values of $U$.
\label{fig:fig3}}
\end{figure}

\textit{Experimental requirements}--%
To gain insight into the experimental parameters, we have numerically calculated $\theta$ as a function of $\varepsilon$ at fixed values of $U$, $t$,
and $\delta h$ by diagonalizing (\ref{Hamiltonian});
the result is shown in Fig.~\ref{fig:fig3}(a).
(Explicit expressions for $J$ and $\tilde{\delta h}$ can be obtained for $|U\pm\varepsilon|\gg t$
by way of a Schrieffer-Wolff transformation \cite{BI}).

Since we assume that $\delta h$ is fixed, the angle $\theta$ will never be exactly $\pi/2$, which is required for perfect Z rotations. The desired values of $t$ and $\delta h$ therefore depend on the error that can be tolerated (see Fig. 3(b)), with $t$ typically exceeding $\delta h$ by more than an order of magnitude. For the X rotations we need $\theta \leq \pi/4$, which gives $\tilde{\delta h} \geq J$, which can be satisfied by moving away from the avoided crossing. The minimum value of $\delta h$ needed for the X rotation is given in Fig.~3(b) for two typical values of $U$.
We note that more detailed calculations including higher orbitals
yield a lower ($<20\%$) value of $J$ \cite{EPAPS}.

In most systems, $t$ can be set by gates or fabrication parameters to anything between $1\,\mu {\rm eV}$ and $1\,{\rm meV}$. Several methods exist for creating an inhomogeneous field $\delta h$: (i) application of an inhomogeneous magnetic field, (ii) different $g$-factors in the two dots--either by composition or confinement~\cite{gfactorcomp}--in combination with a homogeneous magnetic field, and (iii) inhomogeneous nuclear polarizations \cite{Lai2006}. Note that the effect of a fluctuating nuclear field can be diminished by bringing it into an eigenstate~\cite{Nuclearfield}. The electrical bias $\varepsilon$, finally, can be controlled in all quantum dot systems listed in the introduction, by pulsing the source, drain or gate voltage \cite{footnote-t}.

The switching speed of the bias $\varepsilon$ is limited by
adiabaticity constraints:  switching should be sufficiently
fast to guarantee non-adiabatic switching within the qubit 
subspace, $J/\hbar \ll |\dot{\varepsilon}/\varepsilon|$,
but not exceedingly fast, to avoid transitions out of the
computational space to the higher orbital states, such as
the singlet S' in Fig.~\ref{fig:fig1}
($|\dot{\varepsilon}/\varepsilon|\ll U/\hbar$, away from 
the avoided crossing).

\begin{figure}[t]
  \centerline{\includegraphics[width=7cm]{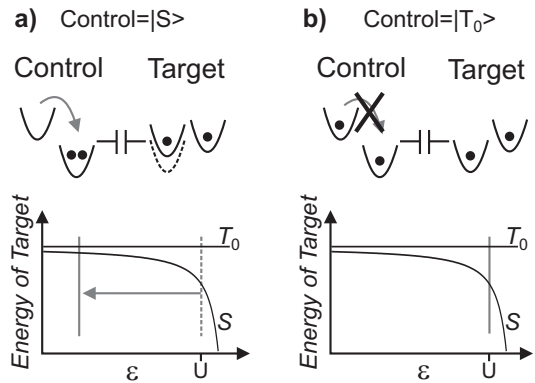}}
\caption{
A CNOT is performed by electrically biasing the 
control qubit, shifting its charge distribution toward the target qubit
(upper panel) \textit{if} it is in a singlet state (a), 
but leaving the charge distribution unchanged 
if it is in the triplet state (b).  
This leads to a conditional shift of the working point of the target 
qubit (lower panels) and to a conditional operation.
\label{fig:fig4}}
\end{figure}

\textit{Controlled-NOT gate}--%
To complete our universal set of 
quantum gates, we require a suitable two-qubit operation, e.g., 
the controlled-NOT (CNOT, or quantum XOR) gate that flips the target qubit 
($|0\rangle \leftrightarrow |1\rangle$)
if the control qubit is in state $|1\rangle$, and otherwise
leaves the target unchanged. This can be achieved by applying
a bias voltage $\varepsilon_{\rm control}$ on the control 
qubit, such that its charge state partly shifts to $(0,2)$
if the qubit state is $\ket{S}$, but remains mostly in 
$(1,1)$ if the state is \ket{T_0} because the $(0,2)$ triplet state is far away in energy~\cite{Petta,Cphase}. Due to the Coulomb 
interaction between the control and the target qubit,
the target qubit will experience a conditional bias shift 
(see Fig.~\ref{fig:fig4}), that can be of the same order as
the interdot Coulomb energy within a single logical qubit.

The CNOT is a conditional $X$-rotation
by  $\gamma=\pi$, thus it is natural to use a
sequence analogous to Eq.~(\ref{x-sequence}),
\begin{equation}
  \label{pre-cnot-sequence}
  \tilde{W}_X(\gamma)=W_\theta(\chi) U_Z(\phi) W_\theta(\chi),
\end{equation}
where $\chi$ and $\phi$ are given in
Eq.~(\ref{chiphi-pi})
in terms of $\theta(\varepsilon)$ and $\gamma$ at the conditional bias
point $\varepsilon$, induced by the charge movement in the
control qubit, and $U_Z$ is the single-qubit $Z$-rotation.
The conditional rotations $W_\theta(\chi)$ about the $\theta$-axis
are analogous to $U_\theta(\chi)$, but instead of being induced by
a direct manipulation of the bias $\varepsilon$, they are
controlled by applying $\varepsilon_{\rm control}$ to
the control qubit, which results in a conditional bias
$\varepsilon$ at the target qubit.

The sequence Eq.~(\ref{pre-cnot-sequence}) for $\gamma=\pi$
is not a true CNOT yet, because (i) the $Z$-rotation $U_Z(\phi)$
is not conditional on the control qubit being in state
$|1\rangle$, but is in fact \textit{always} carried out,
and (ii) the conditional $W_\theta(\chi)$ rotations
also perform a $Z$-rotation in case the control qubit
is in state $|0\rangle$.
In summary, $\tilde{W}_X(\pi)$
does a NOT operation ($X$-rotation by $\pi$) 
on the target if the control is $|1\rangle$
and a $Z$-rotation by $2\chi+\phi$ if the control is
in state $|0\rangle$.  
The true CNOT operation does
\textit{nothing} on the target qubit if the control
is in $|0\rangle$;
it can be obtained with the sequence
\begin{equation}
  \label{cnot-sequence}
  U_{\rm CNOT}=\tilde{W}_X(\pi/2) U_X(\pi) \tilde{W}_X(\pi/2) U_X(\pi),
\end{equation}
canceling the undesired phases if
the control qubit is $|0\rangle$.

In conclusion, we have proposed a universal set of quantum
gates for the $S-T_0$ qubit, consisting of single-qubit rotations about two orthogonal axes
$X$ and $Z$ about arbitrary angles combined with the CNOT gate.
The electrical bias $\varepsilon$
is the only parameter that needs to be tuned fast and with high 
precision, which considerably relaxes the experimental requirements compared
to previous spin-based qubit control proposals and makes our scheme applicable to virtually any quantum dot system.

This research was supported in part by the NSF under Grant No.~PHY99-07949.
We thank the KITP at UCSB, where this work was initiated.
RH is supported by AFOSR, DARPA/MARCO, DARPA/CNID, and ARO;
GB acknowledges funding from the Swiss National Science Foundation and 
NCCR Nanoscience Basel.


\begin{thebibliography}{99}

\bibitem{Petta}
J. R. Petta \textit{et al.}, 
Science {\bf 309}, 2180 (2005).

\bibitem{DelftRO}
R. Hanson \textit{et al.}, Phys.\ Rev.\ Lett.\ {\bf 94}, 196802 (2005).

\bibitem{LD}
D. Loss, D. P. DiVincenzo, Phys.\ Rev.\ A {\bf 57}, 120 (1998).

\bibitem{Levy}
J. Levy, Phys.\ Rev.\ Lett.\ {\bf 89}, 147902 (2002).

\bibitem{WuLidar}
L.-A. Wu and D. A. Lidar,
Phys.\ Rev.\ A {\bf 65}, 042318 (2002);
Phys.\ Rev.\ A {\bf 66}, 062314 (2002).

\bibitem{ByrdLidar}
M. S. Byrd and D. A. Lidar, Phys.\ Rev.\ Lett.\ {\bf 89}, 047901 (2002).

\bibitem{DiVincenzo}
D. P. DiVincenzo, D. Bacon, J. Kempe, G. Burkard, and K. B. Whaley,
Nature {\bf 408}, 339 (2000).

\bibitem{Kyriakidis2005}
J. Kyriakidis and S. J. Penney, Phys.\ Rev.\ B {\bf 71}, 125332 (2005).

\bibitem{Meier}  
F. Meier, J. Levy, and D. Loss, Phys.\ Rev.\ Lett.\ {\bf 90}, 047901 (2003).

\bibitem{Ono2002}
K. Ono, D. G. Austing, Y. Tokura and S. Tarucha, Science {\bf 297}, 1313 (2002). 

\bibitem{SAdots}
M. Bayer \textit{et al.}, Science~\textbf{291}, 451 (2001); B. D. Gerardot~\textit{et al.},
Phys.\ Rev.\ Lett.\ {\bf 95}, 137403 (2005); T.~Ota et al., Phys.\ Rev.\ Lett.\ {\bf 95}, 236801 (2005); E.A. Stinaff~\textit{et al.},  Science \textbf{311}, 636 (2006).

\bibitem{NWs}
M. T. Bj\"{o}rk \textit{et al.}, Nano Lett. \textbf{4}, 1621 (2004).

\bibitem{SiDots}
J. Gorman, D.G. Hasko and D.A. Williams, Phys.\ Rev.\ Lett.\ \textbf{95}, 090502 (2005).

\bibitem{JeroPRB2003}
J. M. Elzerman \textit{et al.}, Phys. Rev. B \textbf{67}, 161308 (2003).

\bibitem{Mason}
N. Mason, M. J. Biercuk, and C. M. Marcus, Science {\bf 303}, 655 (2004).

\bibitem{HayashiPRL2003}
T. Hayashi~\textit{et al.}, Phys.\ Rev.\ Lett.\ \textbf{91}, 226804 (2003).

\bibitem{Taylor}
J. M. Taylor~\textit{et al.}, 
Nature Physics {\bf 1}, 177 (2005);
J. M. Taylor~\textit{et al.}, 
cond-mat/0602470.

\bibitem{footnote1}
Note that $J$ is predicted to go to zero in a finite perpendicular magnetic field \cite{BLD,Kyriakidis2002}, but such a relatively large field also closes the essential single-dot singlet-triplet gap.


\bibitem{BLD}
G. Burkard, D. Loss, and D. P. DiVincenzo, Phys.\ Rev.\ B {\bf 59}, 2070 (1999).

\bibitem{Kyriakidis2002}
J. Kyriakidis \textit{et al.},  Phys.\ Rev.\ B {\bf 66}, 035320 (2002).


\bibitem{footnote-triplets}
We only consider the states with spin projection $S_z\!=\!0$. 
The triplet states $|T_\pm\rangle$ with $S_z\!=\!\pm 1$ can be efficiently decoupled 
from the $S_z\!=\!0$ subspace by chosing $\delta {\bf h}$ purely along $z$ and by 
applying a magnetic field $B_z$ along $z$. 
Note that choosing $z$ in the plane of the dots is preferable to minimize effects 
of $B_z$ on the orbitals.
See \cite{EPAPS} for details. 

\bibitem{EPAPS}
See EPAPS Document No.\ E-PRLTAO-98-036706.
For more information on EPAPS, see \url{http://www.aip.org/pubservs/epaps.html}.


\bibitem{footnote-solution}
The angles $\chi$ and $\phi$ are obtained by solving the transcendental
equations Eq.~(\ref{x-sequence}).

\bibitem{Vandersypen}
L. M. K. Vandersypen and I. L. Chuang, Rev.\ Mod.\ Phys.\ {\bf 76}, 1037 (2005).


\bibitem{HuDasSarmaPRL06}
X. Hu, S. Das Sarma, Phys.\ Rev.\ Lett.\ {\bf 96}, 100501 (2006).

\bibitem{BI}
G. Burkard and A. Imamoglu, Phys.\ Rev.\ B {\bf 74}, 041307(R) (2006).

\bibitem{gfactorcomp}
M.J. Snelling \textit{et al.}, Phys. Rev. B \textbf{44}, 11345 (1991); G. Salis \textit{et al.}, Nature (London) \textbf{414}, 619 (2001); M. T. Bj\"ork \textit{et al.},
Phys. Rev. B \textbf{72}, 201307 (2005).

\bibitem{Lai2006}
C. W. Lai, P. Maletinsky, A. Badolato, and A. Imamoglu,
Phys.\ Rev.\ Lett.\ {\bf 96}, 167403 (2006).

\bibitem{Nuclearfield}
D. Stepanenko, G. Burkard, G. Giedke, A. Imamoglu,
Phys.\ Rev.\ Lett.\ {\bf 96}, 136401 (2006);
D. Klauser, W. A. Coish, D. Loss,
Phys.\ Rev.\ B {\bf 73}, 205302 (2006).

\bibitem{footnote-t}
For some systems, $t$ might have a weak dependence on $\varepsilon$
which could easily be accounted for in Eq.~(\ref{Hamiltonian}).

\bibitem{Cphase}
The controlled-phase gate proposed in Ref.~\cite{Taylor} 
can also be used here, as it does not require 
control over $t$.

\end{thebibliography}
\end{document}